\documentclass{epl}

\newcommand{\vektor}[1]{\mbox{\boldmath $#1$}}
\title{The Non-Abelian Density Matrix Renormalization Group Algorithm}
\author{Ian P. McCulloch and Mikl{\'o}s Gul{\'a}csi}

\institute{Department of Theoretical Physics,
Institute of Advanced Studies, \\
The Australian National University,
Canberra, ACT 0200, Australia}


\begin{document}
\maketitle

\begin{abstract}

We describe here the extension of the density matrix renormalization group
algorithm to the case where the Hamiltonian has a non-Abelian global symmetry
group.  The block states transform as irreducible representations of the
non-Abelian group.  Since the representations are multi-dimensional, a
single block state in the new representation corresponds to multiple
states of the original density matrix renormalization group basis.
We demonstrate the usefulness of the construction via the the
one-dimensional Hubbard model as the symmetry group is enlarged from
$U(1) \times U(1)$, up to $SU(2) \times SU(2)$.

\end{abstract}

In past years, the density matrix renormalization group (DMRG) method
\cite{WhitesOriginal} has been extensively used to study one and two
dimensional strongly correlated electron systems \cite{whitereview}.
This method became very popular when it was realized that it enabled
a level of numerical accuracy for one dimensional systems
that was not possible using other methods \cite{DMR}.

One major drawback of DMRG is that calculations are performed in a
subspace of purely Abelian symmetries, such as the $U(1)$ symmetries of
total particle number
and the z component of the total spin.  Thus one can only obtain a few states in different
total particle number and z component of total spin sectors \cite{noack}.
For models where ferromagnetism emerges the situation worsens, that is,
to determine magnetization, a combination of methods must be employed
which will artificially raise the energy of the higher spin state \cite{ferro}
within the chosen z component total spin sector.

In recognizing the imperative need, to introduce a DMRG method which
has a total spin quantum number naturally implemented, a number of
unsuccessful attempts were previously made (e.g., for the
spin 1 Heisenberg model \cite{rommer} and  t-t'-U model
\cite{daul,sakamoto}).
The most successful previous work on the application of
non-trivial symmetries is the {\em IRF-DMRG} method introduced
by Sierra and Nishino \cite{IRF},
whereby the vertex hamiltonian is first transformed into an interaction
round a face hamiltonian \cite{Baxter}, and then a variant of DMRG
is applied to the IRF model. The IRF model can be chosen
such that it explicitly factors out the global symmetry group. This
technique has been successfully applied to the spin $1/2$ Heisenberg
chain and the XXZ chain with quantum group symmetry $SU_q(2)$ \cite{IRF}
and later, the spin 1 and spin 2 Heisenberg chains \cite{Tatsuaki}.
However, the IRF-DMRG method is complicated by the necessity to
calculate the IRF weights for each interaction term in the Hamiltonian.
The number of non-trivial IRF weights increases rather quickly as the magnitude
of the spins in the system is increased and for a lager global symmetry group.
Thus as far as we know, the IRF-DMRG has not yet been applied to any
more complex models, such as a fermionic system.
In the present work we show that non-Abelian symmetries
can be naturally accommodated into DMRG without the need for a vertex-IRF
transformation. In this form, the starting point of a calculation is
the matrix elements of the single site operators which are relatively
simple to calculate; the number of such elements varies
inversely with the dimension of the irreducible representations of the
global symmetry group, and thus is {\em reduced} for a larger
global symmetry group. For example, all single
site operators of a spin chain are represented as $1 \times 1$ matrices, 
independent of the magnitude of the actual spins. For Hubbard-type models, the number
of single site basis states is reduced from four to two, corresponding to
the spinon and holon states \cite{spinonholon}.

We do not attempt here to give a complete description of the 
DMRG algorithm, instead we refer the
reader to the original description by White \cite{WhitesOriginal} and
more recent reviews \cite{DMR}, and concentrate
on the essential elements of the algorithm that require modification
when using non-Abelian symmetries. These are the construction of
tensor product basis and operators (whether it is through adding
a single site to a block, or joining blocks to construct a superblock),
and the truncation of block states via the reduced density matrix.

We introduce the method by way of the Lie group $SU(2)$.  This symmetry
is readily applicable to all quantum spin systems that can be written
in a form that does not break rotational symmetry. In principle, it is
not difficult to calculate eigenstates of $SU(2)$ for a finite
system by using the Clebsch-Gordan transformation \cite{Biedenharn},
especially in DMRG where the system is built one or two lattice sites
at a time. In this construction, the tensor product of two basis
vectors, labelled here by subscripts 1 and 2, is
\begin{equation}
\vert j m (j_1 j_2 \alpha_1 \alpha_2) \rangle \:
= \: \sum_{m_1,m_2} C^{j_1 j_2 j}_{m_1 m_2 m} \:
\vert j_1 m_1 (\alpha_1) \rangle \: \vert j_2 m_2 (\alpha_2) \rangle \; ,
\label{eq:ClebschGordan}
\end{equation}
where $C^{j_1 j_2 j}_{m_1 m_2 m}$ is the Clebsch-Gordan coefficient. Here
we use the notation $j$ is the total spin quantum number, $S^2 \vert j \rangle
= j(j+1) \vert j \rangle$,
$m$ is the projection of the spin onto the $z-$axis
and $(\alpha)$ is an index that encapsulates the additional labels used in
DMRG (ie, to label the $\alpha$'th basis state of the given quantum numbers).
Bracketed labels are not associated with a quantum number. Constructing
basis states in this way in DMRG suffers from two problems. Applying this
transformation involves two summations for each operator matrix element.
This impacts severely on the computational effort required to construct
the block, and especially the superblock, operators. Secondly.
the direct application of the usual DMRG reduced density
matrix to a wavefunction constructed from some $(jm)$ subspace of
Eq.\ (\ref{eq:ClebschGordan}) does not commute with the $SU(2)$ generators.
Indeed, the wavefunction of an $SU(2)$ invariant system represents the
same physical state independent of the $z$-component of the spin,
so the density matrix of the full system is
\begin{equation}
\rho \: = \: \sum_{m,\alpha^\prime,\alpha} \vert jm(\alpha^{\prime}) \rangle 
		\psi^{}_{jm(\alpha^{\prime})} \psi^*_{jm(\alpha)} 
	\langle jm(\alpha) \vert  \; ,
\end{equation}
where the wavefunction $\vert \Psi_m \rangle = \sum_{\alpha} \psi_{jm(\alpha)}
\vert j m (\alpha) \rangle$ is an eigenstate of total spin, ie
$S^+ \vert \Psi_m \rangle = \sqrt{(j-m)(j+m+1)} \vert \Psi_{m+1} \rangle$.
Using Eq.\ (\ref{eq:ClebschGordan}) and tracing over the right basis, the
$SU(2)$ invariant reduced density matrix for the left block can be constructed,
\begin{equation}
\rho^L_{j_1} (\alpha^\prime_1, \alpha_1) \: = \:
\sum_{m_1,j_2,m_2,\alpha_2}
\psi_{j_1 m_1 (\alpha^\prime_1) ; \: j_2 m_2 (\alpha_2)} \:
\psi^*_{j_1 m_1 (\alpha_1) ; \: j_2 m_2 (\alpha_2)} \; ,
\end{equation}
which acts identically on each $m_1$ component of the basis. This
equation can also be directly by adding an additional constraint to
the reduced density matrix, to force each eigenstate of the reduced
density matrix to be an eigenstate of $S^2$.
For $j=0$ this reduces to the usual DMRG density matrix.
This is seen in conventional DMRG by the
well known $(2j_1+1)$-fold degeneracies in the reduced density matrix
eigenvalues.

Despite the additional overhead of the Clebsch-Gordan transformation, this
construction of $SU(2)$ invariant DMRG works well for small values of $j$,
and is described further in \cite{ajp}.  However, further improvements
are possible. The projection quantum number $m$ can be completely
eliminated using the Wigner-Eckart theorem,
\begin{equation}
\langle j^{\prime} m^{\prime} (\alpha^{\prime})
\vert T^J_M \vert j^{} m^{} (\alpha^{}) \rangle \: = \:
C^{j^{} J j^{\prime}}_{m^{} M m^{\prime}}
\langle j^{\prime} (\alpha^{\prime}) \vert \vert \vektor{T}^J \vert \vert j^{}
(\alpha^{}) \rangle \; ,
\label{eq:WignerEckartTheorem}
\end{equation}
for the $M$'th component of an operator $\vektor{T}^J$ transforming as
a rank $2J+1$ tensor. The quantity $\langle j^{\prime} (\alpha^{\prime})
\vert \vert \vektor{T}^J \vert \vert
j^{} (\alpha^{}) \rangle$ is the {\it reduced matrix element} \cite{Biedenharn}
and is independent of the projection quantum numbers.  This operator can be
considered to act on a reduced basis, given by the complete set of basis
vectors $\vert \vert j (\alpha) \rangle$.  In this form, the superblock
wavefunction for target state $j$ can be written
\begin{equation}
\vert \vert \Psi \rangle \: = \: \sum_{j_{1} j_{2} \alpha_{1} \alpha_{2}}
\psi_{j (j_{1} j_{2} \alpha_{1} \alpha_{2}) } \:
\vert \vert j_{1} (\alpha_{1}) \rangle \: \vert \vert j_{2} (\alpha_{2})
\rangle \;,
\end{equation}
over a product basis given by the Clebsch-Gordan series,
\begin{equation}
\vert \vert j_1 \rangle \otimes \vert \vert j_2 \rangle \: = \:
\vert \vert \; \vert j_1 - j_2 \vert \; \rangle \oplus \cdots \oplus \vert
\vert j_1 + j_2 \rangle \;.
\label{eq:ProductExpansion}
\end{equation}
The reduced density matrix associated with this state is simply
\begin{equation}
\rho^{L}_{j^{}_{1}} (\alpha^{\prime}_{1} \alpha^{}_{1}) \: = \:
\sum_{j_{2}, \alpha_{2}}
\psi^{}_{j^{}_{1} \alpha^{\prime}_{1} ; \: j^{}_{2} \alpha^{}_{2}} \:
\psi^{*}_{j^{}_{1} \alpha^{}_{1} ; \: j^{}_{2} \alpha^{}_{2}} \; .
\end{equation}
The matrix elements of the tensor product of operators
$\vektor{T}^{k_1} \otimes \vektor{U}^{k_2}$
acting on different blocks are given by the Wigner $9j$ coefficients,
\begin{equation}
\begin{array}{c}
\langle j^{\prime}_{} (\alpha^{\prime}_{a} \alpha^{\prime}_{2}
j^{\prime}_{1} j^{\prime}_{2})
\vert \vert \left[ \vektor{T}^{k_1} \otimes \vektor{U}^{k_2} \right]^{k}
\vert \vert j^{}_{} (\alpha^{}_{a} \alpha^{}_{2} j^{}_{1} j^{}_{2}) \rangle
= \\
~ \\
\left[
\begin{array}{ccc}
j^{}_{1}  & j^{}_{2}  & j^{}_{} \\
k^{}_{1}  & k^{}_{2}  & k^{}_{} \\
j^{\prime}_{1} & j^{\prime}_{2} & j^{\prime}_{} \\
\end{array}
\right]
\langle j^{\prime}_{1} (\alpha^\prime_{1}) \vert \vert \vektor{T}^{k_1} \vert
\vert
j^{}_{1} (\alpha^{}_{1}) \rangle \: \langle j^{\prime}_{2} (\alpha^\prime_{2})
\vert \vert \vektor{U}^{k_2} \vert \vert j^{}_{2} (\alpha^{}_{2}) \rangle \; ,
\end{array}
\label{eq:TensorProduct}
\end{equation}
where the $\left [ \cdots \right ]$ term is related to the $9j$ coefficient
\cite{Biedenharn}. With this construction, all steps of the DMRG algorithm
can be performed using only the reduced basis. The importance of this
is that, unlike equation (\ref{eq:ClebschGordan}), there is no
summation involved. The only essential difference from the standard DMRG
formulation is the quantum number dependent $9j$ factor multiplying
each subspace.  Thus, there is no significant computation penalty for
using the $SU(2)$ formulation, as long as the $9j$
coefficients can be calculated efficiently.
In addition, for all two
site interactions, the only two cases that appear are where one
of the block operators in (\ref{eq:TensorProduct}) is the
identity operator, or when block operators are combined to form
a rotational invariant.  In both these cases, the $9j$ coefficient
reduces to a single $6j$ coefficient.

It is worth noting that in the $SU(2)$ formulation, the basis vectors
are exact eigenstates of total spin even after the truncation.  This
is not true, for example, if one attempts to force the ground state
to be in a particular total spin state by adding some suitably
chosen multiple of $S^2$ to the Hamiltonian. Mixing of total spin
states due to numerically near-degenerate states will still occur.
Calculations involving long range
interactions are also affected by the lack of explicit symmetries.
Using a $U(1)$ symmetric basis labelled by the $z$-component of spin only,
interaction terms no longer transform as exact
representations of $SU(2)$ after a truncation.
This can lead to situations where, even
for a large number of kept states, the ground state is a broken
symmetry N\'eel type state \cite{WhiteStripes} and only converges
slowly to an eigenstate of $S^2$. It must be emphasized that this
is purely an artifact of the DMRG algorithm when appropriate
symmetries are not explicitly preserved.

We now have a formulation of DMRG in which the states transform as $2j+1$
dimensional irreducible representations of $SU(2)$. However, it is clear
that the general formulation is essentially independent of the details
of the $SU(2)$ algebra -- given an arbitrary compact global symmetry
group the only modifications to the formulation is a different series
expansion corresponding to Eq.\ (\ref{eq:ProductExpansion}) and coupling
coefficients from Eq.\ (\ref{eq:TensorProduct}). For example, $SO(4)$
is easy to utilize because one can make use of the isomorphism
$SO(4) \simeq SU(2) \times SU(2) / Z^2$, so that the
$6j$ and $9j$ coefficients are simply the product of two $SU(2)$ coefficients.
The component of the algorithm that is model dependent is rather small,
consisting only of the reduced matrix elements of the single site operators,
typically obtained via the Wigner-Eckart theorem.
The $6j$ and $9j$ coupling coefficients of the algebra only appear in the
construction of the blocks, in an identical way all models that admit 
the symmetry group.
This separation of the physical aspects (the reduced matrix elements 
of the single site
operators) and the geometric aspects (the coupling coefficients of 
the symmetry group)
makes the method comparatively easy to apply to a range of models.
This is the main advantage of the non-Abelian formulation over the IRF-DMRG.
In the latter case, without a special effort to factorize the
coupling coefficients, the Boltzmann weights are rather complex quantities, 
especially for large symmetry groups.
We have applied the non-Abelian DMRG successfully to various models with
global symmetries $SU(2)$,
$U(1) \times SU(2)$ \cite{tJ} and $SO(4)$ \cite{Kondo}. The $SU(3)$ case is
in progress.

The computational advantage of the non-Abelian construction is two 
fold: (1) each
reduced basis element corresponds to $2j+1$ basis states of the old
representation, thus the storage requirement for the block operators
is reduced for an equivalent number of block states. (2) the superblock basis 
can be projected onto an exact subspace of arbitrary
total spin. As well as reducing the size of the target
Hilbert space, this greatly simplifies the calculation of excited
states that have total spin less than the total spin of the ground
state. This is very useful for investigating magnetic phase 
transitions \cite{Kondo}.
For ferromagnetic target states (or more generally, target representations
with a dimension greater than one), it is possible to calculate to
first order the splitting of the degenerate states due to a symmetry 
breaking field, trivially in the case of a uniform magnetic field $h$
(where the splitting is just $h m$, for $m = -j,-j+1,\ldots,j$),
or in other cases by calculating the projection of the wavefunction and the 
symmetry breaking operator onto each $z$-component of spin using the Wigner-Eckart
theorem (\ref{eq:WignerEckartTheorem}).

A model for which the non-Abelian formulation is eminently suited is
the Hubbard model \cite{Hubbard},
\begin{equation}
H = -t \sum_{<i,j>,\sigma} \left( c^\dagger_{i,\sigma} c^{}_{j,\sigma} + \mbox{H.c.} \right)
	+ U \sum_i \left( n_{i,\uparrow} - \frac{1}{2} \right) \left( 
n_{i,\downarrow} - \frac{1}{2} \right) \; .
\end{equation}
The main feature of interest in the Hubbard model is the additional
charge $SU(2)$ pseudospin symmetry\cite{pseudospin},
generated by $I^+ = \sum_{i} (-1)^i c^\dagger_{i,\uparrow} 
c^\dagger_{i,\downarrow}$,
$I^- = \sum_{i} (-1)^i c^{}_{i,\downarrow} c^{}_{i,\uparrow}$ and
$I^z = \sum_i \frac{1}{2} (n_{i,\uparrow} + n_{i,\downarrow} - 1)$.
In the resulting reduced $SO(4)$ basis, the Hubbard model
contains only two basis states per site, a spinon of spin 1/2 and 
pseudospin zero,
and a holon of spin zero and pseudospin 1/2. The single site operators are
$2 \times 2$ matrices over this basis.
Table I shows a comparison of the ground state energy for the half-filled
Hubbard model for a 60 site lattice with $t=1$, $U=1$, for
the usual $U(1) \times U(1)$ basis of number of particles and $z$-component of
spin, the $U(1) \times SU(2)$ basis of number of particles and
total spin $S^2 = s(s+1)$, and the $SO(4)$ basis of
total pseudospin $I^2 = i(i+1)$ and total spin.
For the case of half-filling, where the ground state is a spin and
pseudospin singlet, the dimension of the representation $D$ is
equal to the number of basis states that would need to be kept to
achieve the same accuracy using only $U(1)$. Table I shows that
the use of $SO(4)$ symmetry gives an improvement of four orders of
magnitude in the cumulative truncation error and the fractional
error in the ground state energy, for virtually no increase in
CPU time. The main contribution to the variance in the CPU times shown arises from
differences in the number of matrix-vector multiplies being 
performed by the eigensolver, rather than any significant difference in
the CPU time per matrix-vector multiply.

Table II shows a calculation for a higher spin state, at half-filling
with spin $s=5$. In this case, the relative improvement from
using $SU(2)$ symmeteries is not as good.
This has two causes. Firstly, the reduction in the dimension of the Hilbert
space for total spin $s=5$ versus $z$-component of spin $s^z=5$ is not as
big as for the spin zero case. Secondly,
$s=0$ is a special case in which the
number of terms in the Clebsch-Gordan expansion (\ref{eq:ProductExpansion})
for the superblock is exactly one per block quantum number. For higher
spin states, this is no longer true and the number of states in the
superblock progressively increases as the target spin is increased,
as each symmetry sector of the block basis appears multiple times
in the superblock basis. Thus for a fixed number of states
kept, the dimension of the superblock is much larger, with a
corresponding increase in the CPU time.

However we note that targetting a state of spin $j$ is equivalent
to inserting a non-interacting spin of magnitude $j$ into
the system and targetting a state of spin zero of the combined
system + non-interacting spin. Inserting this spin into the
centre of the lattice, in between the system and environment blocks,
is equivalent to targetting the spin $j$ state directly. If
however the non-interacting spin is placed at one end of the chain
and integrated into one of the blocks (it doesn't matter which),
the superblock dimensionality problem is avoided with only very minor
loss of accuracy.
This technique has been used before to target higher spin states
in the IRF-DMRG algorithm \cite{Tatsuaki}. The reflection symmetry
is explicitly broken by the non-interacting spin, however the
increased efficiency is well worth the loss of this symmetry.
During the initial build sweep of the finite size algorithm, 
it is still necessary to target
higher spin states because for a large enough non-interacting spin,
there are not enough spins in the initial four-site block for
the spin zero sector to contain any basis states \cite{Footnote}.
For a target
state of $L$ sites and spin $j$, the actual target state
during the build sweep when the lattice size is $l$ sites
(not counting the non-interacting spin),
is $j(1-l/L)$, rounded to the nearest permissible half-integer.
Table III shows the improvement when this form of targetting is used.
As this table shows, the additional superblock states that are included
when the direct targetting method is used have negligible effect on
the variational energy.

We have extended the DMRG algorithm so that the block
and superblock basis states transform as representations of
an arbitrary compact non-Abelian global symmetry group, and demonstrated
the improvement in accuracy for the Hubbard model utilizing
spin and pseudospin symmetry. This is
a true generalization of the conventional DMRG algorithm in that,
if we instead use the coupling coefficients of $U(1)$
instead of $SU(2)$ in Eq. (\ref{eq:TensorProduct}),
the original DMRG algorithm is recovered exactly. Thus optimizations
such as efficiently storing the block operators \cite{Peter}, and transforming
the obtained wavefunction to be the initial vector for the
next DMRG iteration \cite{WavefunctionAcceleration} apply to the non-Abelian
case in a straight forward manner and were used in the current calculation.
We have shown that, for the ground state of the half-filled Hubbard model,
keeping only 300 $SO(4)$ states
is equivalent to keeping over 1700 states of the $U(1)\times U(1)$ basis
of the original DMRG formulation. As the spin of the target state is increased,
the accuracy improvement diminishes because the difference between
the dimension of the Hilbert space of the
total spin symmetry sector and the highest weight $z-$component of spin sector
is reduced. Directly targetting a higher spin state is inefficient because
the Clebsch-Gordan expansion implies that the dimension of the superblock will
be much larger for the same number of block states. This inefficiency can be
easily overcome by using a non-interacting spin to force the target state
into the singlet symmetry sector.

\acknowledgments

One of the authors (IPM) wishes to thank Los
Alamos National Laboratory for hospitality, where some parts
of this work were finalized. We also thank Tomotoshi Nishino for
useful discussions on the IRF-DMRG algorithm.
All calculations were performed on a desktop computer with
a 500 MHz Athlon processor.

\begin{table}
\caption{Comparison of $U(1) \times U(1)$, $U(1) \times SU(2)$
and $SO(4)$ basis for the groundstate of the half-filled Hubbard
model for a 60 site lattice, at $t=U=1$. Number of states kept $m$,
dimension of the group representation $D$, energy $E$, fractional
error in the energy, cumulative truncation error over the sweep
$1-\sigma$, CPU time in seconds per sweep.}
\[
\begin{array}{lrr r@{.}l r@{\times}l r@{\times}l  r}
\hline \hline
\mbox{basis} & \multicolumn{1}{c}{m} & \multicolumn{1}{c}{D} &
\multicolumn{2}{c}{E} & \multicolumn{2}{c}{(E - E_g) / |E_g|} &
\multicolumn{2}{c}{1-\sigma} & \mbox{CPU} \\ \hline
U(1)\times U(1)   & 100 & 200  & -61 & 7484986435 & 5.2 & 10^{-5}  & 
5.3 & 10^{-4} &  10        \\
U(1)\times U(1)   & 200 & 200  & -61 & 7514641444 & 4.5 & 10^{-6}  & 
4.8 & 10^{-5} &  41        \\
U(1)\times U(1)   & 300 & 300  & -61 & 7516910404 & 7.9 & 10^{-7}  & 
8.8 & 10^{-6} & 110        \\
U(1) \times SU(2) & 100 & 226  & -61 & 7515581914 & 2.9 & 10^{-6}  & 
3.1 & 10^{-5} &  15        \\
U(1) \times SU(2) & 200 & 468  & -61 & 7517319907 & 1.3 & 10^{-7}  & 
1.4 & 10^{-6} &  64        \\
U(1) \times SU(2) & 300 & 716  & -61 & 7517389831 & 1.4 & 10^{-8}  & 
1.5 & 10^{-7} & 158        \\
SO(4)             & 100 & 526  & -61 & 7517351742 & 7.6 & 10^{-8}  & 
8.4 & 10^{-7} &  18        \\
SO(4)             & 200 & 1136 & -61 & 7517397636 & 1.4 & 10^{-9}  & 1.5 
& 10^{-8} &  71        \\
SO(4)             & 300 & 1766 & -61 & 7517398448 & 9.9 & 10^{-11} & 
1.0 & 10^{-9} & 133        \\
\hline \hline
\end{array}
\]
\end{table}

\begin{table}
\caption{Comparison of the $U(1) \times U(1)$, $U(1) \times SU(2)$
and $SO(4)$ basis for the lowest spin 5 excited state for the
half-filled Hubbard model on a 60 site lattice with $t=U=1$.}
\[
\begin{array}{lr r@{.}l r@{\times}l r@{\times}l  r}
\hline \hline
\mbox{basis} & \multicolumn{1}{c}{m} &
\multicolumn{2}{c}{E} & \multicolumn{2}{c}{(E - E_g) / |E_g|} &
\multicolumn{2}{c}{1-\sigma} & \mbox{CPU} \\ \hline
U(1)\times U(1)  & 100 & -59 & 5701792131 & 4.0 & 10^{-5} & 3.9 & 
10^{-4} &  11        \\
U(1)\times U(1)  & 200 & -59 & 5723270633 & 3.6 & 10^{-6} & 3.8 & 
10^{-6} &  41        \\
U(1)\times U(1)  & 300 & -59 & 5725015232 & 6.3 & 10^{-7} & 6.8 & 
10^{-5} & 102        \\
U(1)\times SU(2) & 100 & -59 & 5702795890 & 3.8 & 10^{-5} & 3.9 & 
10^{-4} &  26         \\
U(1)\times SU(2) & 200 & -59 & 5723402180 & 3.3 & 10^{-6} & 3.7 & 
10^{-5} &  90        \\
U(1)\times SU(2) & 300 & -59 & 5725035338 & 5.9 & 10^{-7} & 6.8 & 
10^{-6} & 207        \\
SO(4)            & 100 & -59 & 5723565660 & 3.1 & 10^{-5} & 3.5 & 
10^{-5} &  32        \\
SO(4)            & 200 & -59 & 5725315037 & 1.3 & 10^{-7} & 1.4 & 
10^{-6} & 127       \\
SO(4)            & 300 & -59 & 5725381508 & 1.4 & 10^{-8} & 1.5 & 
10^{-7} & 297        \\
\hline \hline
\end{array}
\]
\end{table}

\begin{table}
\caption{Comparison of the $U(1) \times U(1)$, $U(1) \times SU(2)$
and $SO(4)$ basis for the lowest spin 5 excited state for the
half-filled Hubbard model on a 60 site lattice with $t=U=1$,
using a non-interacting spin to target the appropriate symmetry
sector.}
\[
\begin{array}{lr r@{.}l r@{\times}l r@{\times}l  r}
\hline \hline
\mbox{basis} & \multicolumn{1}{c}{m} &
\multicolumn{2}{c}{E} &\multicolumn{2}{c}{(E - E_g) / |E_g|} &
\multicolumn{2}{c}{1-\sigma} & \mbox{CPU} \\ \hline
U(1)\times SU(2) & 100 & -59 & 5702385716 & 3.9 & 10^{-5} & 3.8 & 
10^{-4} &  14         \\
U(1)\times SU(2) & 200 & -59 & 5723344203 & 3.4 & 10^{-6} & 3.6 & 
10^{-6} &  47        \\
U(1)\times SU(2) & 300 & -59 & 5725027479 & 6.1 & 10^{-7} & 6.7 & 
10^{-6} & 113        \\
SO(4)            & 100 & -59 & 5723497975 & 3.2 & 10^{-6} & 3.4 & 
10^{-5} &  16        \\
SO(4)            & 200 & -59 & 5725312667 & 1.3 & 10^{-7} & 1.4 & 
10^{-6} &  64       \\
SO(4)            & 300 & -59 & 5725381253 & 1.4 & 10^{-8} & 2.2 & 
10^{-7} & 148        \\
\hline \hline
\end{array}
\]
\end{table}


\begin{thebibliography}{0}
\bibitem{Footnote}This problem has so far constrained IRF-DMRG calculations to
the calculation of low lying spin states only.
\end{thebibliography}
\end{document}